\documentstyle[aps,prc,preprint,tighten]{revtex}

\font \msam=msam10 
 
\begin{document} 
\draft
\title {\bf Improved $d$+$^4$He potentials by inversion, 
the tensor force and validity of the double folding model.}

\author {V.I.Kukulin, V.N.Pomerantsev}
\address{Institute of Nuclear Physics, Moscow State University,\\ 
Moscow 119899, Russia.}
\author {S.G.Cooper}
\address{Physics department, The Open University, Milton Keynes \\ 
MK7 6AA, UK.}
\author {S.B.Dubovichenko}
\address{Dept. of Physics, Kazakh State Iniv., Almaty. \\} 
 
\date{\today } 
\maketitle 
 
\begin{abstract} 
Improved potential solutions are presented for the inverse scattering
problem for $d$+$^4$He data. The input for the inversions includes
both the data of recent phase shift analyses and phase shifts from RGM
coupled-channel calculations based on the NN Minnesota force. The
combined calculations provide a more reliable estimate of the odd-even
splitting of the potentials than previously found, suggesting a rather
moderate role for this splitting in deuteron-nucleus scattering
generally. The approximate parity-independence of the deuteron optical
potentials is shown to arise from the nontrivial interference between
antisymmetrization and channel coupling to the deuteron breakup
channels. A further comparison of the empirical potentials established
here and the double folding potential derived from the M3Y effective
NN force (with the appropriate normalisation factor) reveals strong
similarities.  This result supports the  application of the
double folding model, combined with a small Majorana component, to the
description even of such a loosely bound projectile as the deuteron.
In turn, support is given for the application of
iterative-perturbative inversion in combination with the double
folding model to study fine details of the nucleus-nucleus
potential. A $d$-$^4$He tensor potential is also derived to reproduce
correctly the negative $^6$Li quadrupole moment and the $D$-state
asymptotic constant.
\end{abstract} 
 
\newpage 
\centerline{Contents of the paper.} 
\begin{enumerate} 
\item Introduction 
\item The role of parity dependence in IP inversion
\item Stabilised inversion of the  $d$-$^4$He empirical data for  
$0<E_d\le $15 MeV
\item Inversion of RGM  $S(l)$ with breakup channel  
contributions
\item Validity of double folding model for $d$-$^4$He interaction and 
similar systems. 
\item Tensor $d$-$^4$He potential and the $^6Li$-quadrupole moment. 
\item Conclusions 
\end{enumerate} 
 
\section{Introduction} 
\label{intro} 
Many basic features of the interactions between  light composite
particle are now well established. In particular, a good description
\cite{bib1}-\cite{bib30}, of the interactions of light nuclei such 
as $d+t$, $t $+$^3$He, $^4$He+$^4$He etc. is obtained from a deep
attractive potential with Pauli-forbidden states with the addition of
parity dependence, Young scheme splitting and a spin-orbit
interaction.  Importantly, this type of interaction is now justified
from microscopic considerations, both in quasi-classic picture
\cite{bib3} and also in quantum mechanical shell-model framework
\cite{bib1,bib7}.  These results lead to a general comprehensive
understanding of the relations between various interaction models
which appear, at a first glance, to contradict to each other.
 
Nevertheless, a number of finer features of the interaction, notably
the role of the dynamic polarisation of loosely bound projectiles such as
$d$ or $^{6,7}$Li when combined with antisymmetrization effects, are
not yet fully understood despite many previous efforts. The existing
problems and contradictions are illustrated by the following example.
On the one hand, it is well known (see
e.g. \cite{bib1},\cite{feschb}-\cite{bib13}) that antisymmetrization
and coupled-channel effects in composite particle scattering
unavoidably result in complex nonlocal and energy-dependent
potentials, in which the breakup channels (with their various final
angular momenta) lead to rather peculiar contributions in $r$-space
and the specific energy dependence. On the other hand, for a long time
it has been standard practice to describe the elastic scattering of
deuterons and nuclei such as $^6$Li by the standard optical model (via
local potentials) with a smooth energy dependence, or even by global
optical models \cite{bib14}.
 
The low energy $N+^4$He interaction presents a further example of
these problems.  An attempt undertaken long ago by Satchler {\em et al},
\cite{bib15}, to describe $N+^4$He phase shifts for $E_N<$20 MeV by a
standard optical potential of Woods-Saxon form, required a
significantly energy dependence in the {\em radial form} of the
potential (i.e. of geometric parameters).  Further studies based on
$S$-matrix to potential inversion, \cite{bib16,pa-emp}, showed that
the $N+^4$He phase shifts can be described excellently over a wide
energy range ($E_N<$65 MeV) with a gaussian-like potential with
odd-even splitting and with a small energy dependence in the potential
depth alone. This odd-even splitting effect is a direct consequence of
antisymmetrization, \cite{bib1,rgm-rep,pa-rgm}.  Thus the artificial
energy dependence found by Satchler {\em et al} \cite{bib15} reflects only
the effect of omitting the odd-even splitting in the $N+^4$He
potential.
 
Similar problems are also expected to arise for $d$+$^4$He and
analogous systems. Moreover, many previous studies of deuteron
scattering have demonstrated \cite{bib12,bib13,im-break} the strong
contribution of deuteron breakup channels, in particular with $\Delta
l$=2, i.e. through the excitation of the $D$-state in N-N subsystem.
 
We can then ask, in what way can the above strong nonlocal
contribution be incorporated into, say, the double folding potential,
which, from a first glance, contains no such effects?  (In this
connection we must emphasise that the usual type of nonlocalities
considered in literature \cite{bib8,bib9} differs from the
energy-singular terms arising when the virtual breakup channels are
excluded).  An important problem then arises to formulate the above
nonlocal and specific energy dependent effects into a language of
``exact'' optical potentials. By this term we do not mean a standard
phenomenological optical potential with a prescribed form (and fitted
parameters) but instead some ``exact'' potential which is
reconstructed by inversion from the data or from microscopic theory.
 
Our plan of study is as follows. As a first approach iterative
perturbative (IP) inversions are presented, based on phase-shift
analyses of experimental data.  These inversions follow
Ref.~\cite{bib16}, in directly calculating an explicit Majorana
exchange force, instead of applying separate inversions for even and
odd partial amplitudes as in previous determinations of the empirical
$d$ + $\alpha$ potential, \cite{bib20}.  In this way the even and odd
partial amplitudes are treated simultaneously and the inversion
process is considerably stabilised.  In the second approach, an
extended RGM-study is presented for the same system in order to gain a
deeper understanding of the complicated interplay between
antisymmetrization and projectile breakup effects. Increasing the
number of coupled channels in these RGM calculations leads to three
sets of microscopic models for the $d-^4$He interaction:
\begin{enumerate} 
\item[-] the direct $d-^4$He potential without incorporation of
any exchange effects and which is related to standard double
folding (DF) models (see e.g.~\cite{m3y-rep});
\item[-] the one channel ($d+^4$He) RGM model with full 
antisymmetrization but  with a ``frozen'' deuteron and
\item[-] the multichannel RGM model inclusive of breakup channels. 
\end{enumerate} 
RGM potentials are determined by inversion from the phase shifts 
of the last two models. The comparison of these different 
potential models then yields the contributions of antisymmetrization 
and deuteron breakup effects in terms of the local potential. Our 
approach has one further advantage, since the considerable flexibility 
of the inversion method allows the determination of potentials, which 
describe the system under study not at a single energy but for a wide 
energy range simultaneously (previously denoted as ``mixed case'' inversion 
\cite{pa-emp,bib17,bib18}).  The further comparison of our new parity 
dependent solutions obtained from the empirical data set and the RGM 
phase shifts allows {\em quantitative } deductions to be made on the 
validity of the DF model for the deuteron projectile (as applied in
\cite{bib10} for example). 
 
The final part of our study is dedicated to a reconstruction of the 
tensor part of the $d$-$^4$He interaction. Here we study the range and 
depth of the $d$-$^4$He tensor force and compare it with the same 
quantities for central and spin-orbital components of the same system. 
 
The contents of this work are then as follows. In
Section~\ref{invs-sect} we describe, in brief, our inversion method
and discuss the importance of including odd and even partial waves
simultaneously in the inversion. Section~\ref{emp-sect} describes 
the empirical input data used and presents a new solution for the
empirical case.  Section~\ref{rgm-sect} is devoted to inversion of RGM
calculations for $d$+$^4$He, which include coupling to the virtual
breakup channels. Section~\ref{dfm-sect} contains a detailed
discussion of validity of the DF model. In Section~\ref{tens-sect}, we
describe the calculation of a tensor $d$-$^4$He potential which nicely
fits the $^6$Li quadrupole moment and the tensor mixing parameter
$\varepsilon _1$. Our findings are summarised in a concluding section.

\section{The role of parity dependence in IP inversion} 
\label{invs-sect} 
 
The inversion method used in the next sections has been developed in
Moscow (in the Moscow State University) 
\cite{bib20,bib17,bib18,bib19} and independently in the Open University by
Mackintosh and coworkers, (originally in Ref.~\cite{mk-break},
with further developments and references in Ref.~\cite{bib16}).  It
has been named as the iterative-perturbative (IP) method by the latter
group and as the linearised iterative method by the Moscow group. The
two approaches have only minor differences which relate to the choice
of inversion basis and details of the iteration process. The overall
method is extremely flexible and convenient to use. It enables us to
reconstruct many types of potentials, central, spin-orbit and tensor,
real and complex, both from phase shifts or directly from scattering
observables.  The appropriate input quantities, e.g. $S$-matrix,
differential cross sections, analysing powers etc.  are nonlinear
functions of the interaction potential and so may be considered as a
functional response to variations in the interaction potential. In
general, excluding some special cases, a small variation of the
interaction potential is directly related to a small variation of the
$S$-matrix or other response function.
 
The basic idea behind the approach is then a local linearization of 
the response function in neighbourhood of a given point in an 
appropriate functional space. This local linear approximation, 
together with an expansion of the unknown potential in some complete 
basis (for example the Fourier expansion in an orthogonal basis, 
\cite{bib20,bib17,bib18,bib19} although gaussian functions 
worked well in inversions for $p$ + $\alpha$, \cite{bib16}), results
in a set of linear algebraic equations to be solved to give the
expansion coefficients.  The initial approximation (i.e. initial
values of the expansion coefficients or a starting reference potential) is
usually chosen based on physical considerations. In particular, as
will be shown in later sections, the results of the present paper
suggest that a good choice for the initial potential would be the
DF potential.  The initial potential
is then corrected with step-by-step iterations. The iteration
procedure converges rapidly (5 -- 7 iterations at most are usually
required) and gives a stable solution for the first expansion
coefficients of the sought potential, where the number of
reconstructed coefficients depends directly on the completeness and
consistency of the input data.
 
The approach developed by the Moscow group has also been successfully applied, 
\cite{bib21} to reconstruct interactions in the field of heavy ion scattering  
in systems such as $^{12}$C$+^{13}$C, $^{16}$O$+^{17}$O etc. The 
algorithm developed at the Open University has also been applied to a 
wide range of systems; for example to $p$ + $^{16}$O directly from 
experimental observables, \cite{p-ox16}, to mixed case $N$-nucleus phase 
shifts calculated from non-local interactions, \cite{pb-nnuc}, and to 
single energy empirical $S$-matrices, for nuclei such as 
$^{16}$O$+^{16}$O, \cite{o16-o16}, and for $^{11}$Li scattering, 
\cite{11li}. 

 \subsection{The role of odd-even potential splitting.}
 
The presence of well expressed low-energy resonances in the $L$=2
partial waves and near the $S$-matrix pole in the $S$-wave has been
shown previously \cite{bib20,bib18,bib19} to render the solution of
the inverse problem for the above channels so reliable that the data
for even a small energy interval 0-5 MeV is sufficient to give quite a
correct reconstruction of $d$-$^4$He interaction potential in even
partial waves.  In sharp contrast to this, the odd partial amplitudes
are derived from the phase shift analysis with large errors due to a
low sensitivity of majority of observables (cross sections and tensor
analysing powers) to these odd-parity partial phase shifts.
As a result, the odd parity solution of the inversion has large
errors which become more enhanced at the higher energies
\cite{bib20,bib19}. Hence, when the odd and even partial amplitudes are 
treated {\em independently} in the inversion \cite{bib20,bib18,bib19},
the magnitude and radial form of the Majorana interaction term cannot
be established with sufficient accuracy. However, the structure of
this term is very important for our understanding of the role and
significance of exchange effects in nucleus-nucleus interactions.
 
The general form of operator (or potential) of particle  
interaction can be expressed as follows: 
\begin{equation} 
 V=V_W+\hat {M}V_M   \label{def_v}
\end{equation} 
where $V_W$ is the Wigner type interaction which includes the central,  
spin-orbit and tensor interaction terms,  i.e., 
\begin{equation} 
V_W=V^{(c)}_W+V^{(sl)}_W \mbox{${\bf l}$}\mbox{\boldmath $\cdot$}
\mbox{${\bf s} $}  +V^{(t)}_W\hat {S}_{12}, 
\end{equation} 
$V_M$ is the Majorana type interaction which includes, in general,  
similar terms, i.e. central, spin-orbit and tensor, and $\hat {M}$ is the  
Majorana exchange operator.  In the case here, for the 
interaction of two nuclei, $a$  
and $b$, $\hat {M}=P_{ab}=(-1)^{l_{ab}}$ and $l_{ab}$ is a relative angular  
momentum of the pair $a$ and $b$. Furthermore, in the standard approach the  
operator (1) can be rewritten in the form: 
\begin{eqnarray}V & = & \left (V^{(c)}_W+P_{ab}V^{(c)}_M\right )+\left  
(V^{(ls)}_W+P_{ab}V^{(ls)}_M\right )\mbox{${\bf l}$}\mbox{\boldmath $\cdot$}
\mbox{${\bf s} $}  
+\left (V^{(t)}_W+P_{ab}V^{(t)}_M\right )\hat{S}_{12} \nonumber 
\\ 
& = & \sum _{k=c,ls,t}\left (V^{(k)}_W\pm V^{(k)}_M\right )A^{(k)} 
\end{eqnarray} 
where the operators 
\begin{equation} 
A^{(k)}=\left \{ \begin{array}{lr} 
                               1, & k=c\\ 
                              \mbox{${\bf l}$}\mbox{\boldmath $\cdot$}
\mbox{${\bf s} $}, & k=ls\\ 
                               \hat {S}_{12}, & k=t \footnotemark

\end{array}\right.
\label{def_ak} 
\end{equation} 
\footnotetext{In
Sect.~\ref{tens-sect} we use the standard  ${S}_{12}$  form for the
tensor operator. A full classification for the tensor force based on 
symmetry principles was given by Satchler, \cite{bib9}.
See also Robson, \cite{robson}}                     
and the plus sign relates to even partial waves and minus sign to odd  
partial waves. Now the interaction in the odd and even partial waves are  
uncoupled and each one can be parametrised and determined {\em  
independently} from the other\footnote {Formal mixing of odd and  
even partial waves is excluded due to parity conservation in nuclear  
interaction. Here we neglect a very small degree of mixing of the odd and  
even components due to the parity nonconserving terms.}.  
 
The procedure by which two inversions are applied to establish  
the odd and even 
components of interaction potentials will be 
denoted here as ``{\em 
separate}'' inversion. This method has been previously applied to 
  $p+^4$He \cite{pa-emp} and 
$d$+$^4$He \cite{bib20,bib19}.  However, as discussed above,
the separate determination of 
even and odd potential components induces significant errors in the 
appropriate potential if one set of partial amplitudes, viz. odd or 
even, with $J=L+1$ or $J=L-1$ etc. is determined from the phase shift 
analysis with considerable errors and this is quite often the situation. 
 
The alternative approach is to derive the odd and even potential
components simultaneously using the original form of interaction
operator (3), i.e. to be denoted ``{\em simultaneous}'' inversion, and
this procedure was first introduced in Ref.~\cite{bib16}. For the
inversion of $p+^4$He empirical phase shifts the two approaches lead
to very similar potentials. However simultaneous inversion was
necessary to provide a stable parity-dependent inversion of $p+^{16}$O
scattering observables, \cite{p-ox16}, because the Majorana term is
very small in comparison to the Wigner type interaction.  On the same
note, it is common practice to apply inversion to all members of
the families of partial amplitudes, $J=L+1$ and $J=L-1$
simultaneously, i.e. to determine central and spin-orbit potentials
directly rather than to calculate potentials for the separate spin
channels with $J=L-1$ or $J=L+1$ and so on.

The general philosophy behind the simultaneous inversion approach is
the following. While the Majorana components describe exchange
terms in the nucleus-nucleus interaction, \cite{rgm-rep,bib12}, these
components represent only a minor effect of the exchange correction
compared to that contributing to the main Wigner term, which appears
in the conventional simple and double folding models for
nucleus-nucleus interactions. This difference suggests that a more
stabilised inversion is obtained by expanding $V_W$ and $V_M$ in
separate basis sets, as in Ref.~\cite{p-ox16}, with different scale
parameters in each case, i.e.
\begin{equation} 
V_W(r)=\sum^{N_1}_{i=0}C^{(1)}_i\phi ^{(1)}_i(r) 
\end{equation} 
\begin{equation} 
V_M(r)=\sum^{N_2}_{j=0}C^{(2)}_j\phi ^{(2)}_j(r) 
\end{equation} 
Here $\phi ^{(1)}$ and $\phi ^{(2)}_j$ are basis sets with appropriate 
radial scale parameters.  
 
This approach then leads to more reliable determination of the radial
form and range of $V_W$ and $V_M$ compared to the potentials resulting
from the separate inversion which may have significant uncertainties
in the Majorana component.  The inversion of empirical $d$+$^4$He
phase shifts, described further in the following section, well
illustrates the importance of simultaneous inversion.  The improvement
arises because both components are simultaneously determined from even
(well determined) and odd (poorly determined) partial amplitudes on
the equal footing. If one then combines such simultaneous
determination with a step-by-step extension of the basis in (5) and
(6), which is equivalent to the demand of maximal smoothness for each
component, $V_W$ and $V_M$, the inversion is stabilised.

\section{Stabilised inversion of  the  $d$-$^4$He empirical data for  
$0<E_d\le $15 MeV} 
\label{emp-sect} 
\subsection{The empirical $d$+$^4$He phase shifts} 
\label{dhe-ps} 
 
In the next section we reconstruct the $d$-$^4$He interaction
potential from phase-shift  analysis (PSA) data using the
algorithm developed by the Moscow group.  The inversions are based on
two separate sets of phase-shift data:
\begin{enumerate} 
\item[(i)] single energy PSAs of the Z\"{u}rich group \cite{bib25}  
established for the energy  
range  6 - 43 MeV and 
\item[(ii)] the extended energy-dependent PSA of Kuznetsova {\em et al}
\cite{bib26} for 0 - 10 MeV, with a special emphasis on the  
lowest energy region. 
\end{enumerate} 

As in previous inversions based in this PSA, tensor potential
terms are excluded from all the inversion calculations presented here
except those
described in Sect.~\ref{tens-sect}, since the available data on the
mixing parameters are too unreliable to produce a stable inversion.
This problem arises due to the very weak influence of the tensor
interaction on the low-energy phase-shifts, e.g. $\varepsilon_1$
differs from zero only in the narrow region near 6.5 MeV, where the
$^3S_1$ and $^3D_1$ phase shifts are close in value. For similar
reasons, only real potentials are determined due to the limitations in
the inelasticity parameters. However, in Sect.~\ref{tens-sect} we
present a  reconstruction of the $^4$He + d tensor potential from the $^6$Li
quadrupole moment and the existing PSA data for the $\varepsilon_1$
mixing parameter.

Although the PSA-solutions of Z\"{u}rich group contain many
irregularities and some non-monotonic behaviour, especially for $E_d>
20$ MeV, the phase shifts in the energy range 6 - 14 MeV \cite{bib25}
are satisfactorily smooth functions of energy. On the other hand, the
energy-dependent PSA \cite{bib26} gives very reliable $S$- and
$D$-wave phase shifts for $E_d< 6$ MeV, which strongly stabilises the
inversion for both even and odd partial waves (see
Sect.~\ref{invs-sect}).  Therefore in this work we present results for
inversions based on the phase shifts at energies up to 15 MeV only.  A
reconstruction of the $d$-$^4$He interaction using all the Z\"{u}rich
PSA data (up to 43 MeV), but by separate inversion, has been presented
previously, \cite{bib20}.

\subsection{Inversion from the $d$+$^4$He phase shift analysis data.} 
 
In this section we compare two solutions of the $d$+$^4$He inverse problem, 
for the  energy range  
$0<E_d\le $15 MeV, obtained as follows: 
\begin{enumerate} 
\item [(i)] by separate inversion of odd and even partial amplitudes,  
\cite{bib20},  
\item[(ii)]  by simultaneous inversion of  both amplitudes (present work).  
\end{enumerate} 
As a starting approximation we take only one term in the expansions of
each potential component, e.g. central, spin-orbit etc. Following our
initial suggestions, \cite{bib17} we have used a s-wave orthonormal
harmonic oscillator basis, for which the first term of the series
corresponds simply to a gaussian. Then, in each subsequent stage, we
add one term to each of the above components (either to each component
alternately or to all components at once). Such a procedure converges
very rapidly and enables us to reliably control the process. If the
input information is insufficient to determine the next expansion
term, either the iterations fail to converge or the uncertainties in
the expansion coefficients become too large, $\sim 100$\%,
\cite{bib17}. The convergence of the method had been investigated  
previously \cite{bib17,bib18} and here we only give the final results
of the inversion.

 In Figs.\ref{fig-1ab}a, \ref{fig-1ab}b and  
\ref{fig-2ab}a, \ref{fig-2ab}b we compare the potentials found  
for these two cases (the central components are displayed on 
Fig.~\ref{fig-1ab} while the spin-orbit components are shown on  
Fig.~\ref{fig-2ab}). 
 
The degree to which these two potentials reproduce the phase shifts is
shown on Figs.~\ref{fig-3} and~\ref{fig-4} for the even and odd
$l$-values respectively.  The apparent disagreement between the
predictions of the inversion and the data of the phase shift analyses
for the $P$-waves, (Fig.~\ref{fig-4}), probably arises due to the
large errors in the PSA data for these amplitudes,
\cite{bib25,bib26}. In particular, the results of accurate Faddeev
calculations for $d$-$^4$He scattering
\cite{bib7,bib31} show a much better agreement with our predictions 
than with the PSA.  So our simultaneous 
inversion for the $P$-waves may be  closer to the ``true'' solution than is 
apparent from Fig.~\ref{fig-4}. 
 
It is evident from Figs.~\ref{fig-1ab} and~\ref{fig-2ab} that the 
modifications to the potentials found when switching from the separate 
inversion to the simultaneous approach are rather small only for the even 
central potential, and are quite significant for the even 
spin-orbit and all the odd components. This improvement is due to the
stabilisation of the simultaneous inversion.  
 
Strong changes are also found in the odd potential on inclusion of the 
$F$ phase shifts in the inversion (the dashed curve  in 
Fig.~\ref{fig-1ab}b), when compared with inversion based on the $P$ 
wave amplitudes alone (the dotted curve in Fig.~\ref{fig-1ab}b). 
These effects arise despite  the large errors in the $F$-phase 
shifts of the PSA 
\cite{bib25}.  Clearly, the incorporation of such additional and 
{\em independent} input data for the inversion results in a further strong 
stabilisation of the inversion. 
 
The above conclusion is in agreement with our initial general 
expectation that incorporation even of rather ``noisy'' but {\em 
independent} input data into the inversion, together with 
some definite additional constraint (e.g. requiring some degree of 
smoothness in the potentials), leads to a noticeable stabilisation of 
the solutions. 
This general feature of the IP inversion procedure then convincingly 
distinguishes our approach from more traditional and strict methods 
like Gel'fand-Levitan or Marchenko approaches. In the applications of 
the latter procedures, it is impossible to incorporate essentially 
incomplete or ``noisy'' data into the initial data set. 
 
Turning now to the spin-orbit potential, we find a further
illustration of the above conclusions. In fact, our results show that
the Majorana component of spin-orbit interaction cannot be reliably
derived from the existing PSA data. The reasons for this are partly
due to the errors in the PSA data, but also due to the small size of
the Majorana spin-orbit term.
 
This small Majorana spin-orbit term is justified by a comparison of
the Wigner and Majorana components for central potential (see
Fig.~\ref{fig-5}).  Both the inversions based on RGM-calculations (see
Fig.~\ref{fig-2coop} in Section~\ref{rgm-sect}) and those based on the
PSA data lead to central interactions for which the strength of
Majorana component is smaller than the Wigner term by an order of
magnitude.  This result confirms the general expectations that the
role of the Majorana term is then as a minor correction to the main
Wigner term. Evidently, if a similar relation between Wigner and
Majorana terms exists for the spin-orbit interaction, there is no
chance of determining a Majorana (i.e.  exchange) spin-orbit term from
the existing PSA data.
 
Thus, summarising our findings in this Section, we conclude that the  
effects of the odd-even splitting for $d + ^4$He are rather moderate. 
The significance of the effects are also  illustrated by the values of the  
volume integrals, 
\begin{equation} 
J^{(k)}=-\frac{4\pi }{A_T A_P}\int V^{(k)}(r)r^2dr 
\end{equation} 
for the different components $V^{(k)}$ of the inverted potential, 
Eqs.~\ref{def_v}-\ref{def_ak}.
Table~\ref{invs-cso} displays the magnitudes of volume 
integrals $J_k$, for the potentials reported above and these values  
corroborate the above conclusions.  

We have also estimated the value of $J^{(c)}$ for the Majorana terms
of the potentials determined by simultaneous inversion. 
These potentials give, 
$$J^{(c)}_{even}-J^{(c)}_{odd}=2J^{(c)}_M \simeq 0.17 J^{(c)}_W,$$ 
i.e. $J_M$ is about 8.5 $\%$ of $J_W$.  Hence, as was expected 
beforehand, the Majorana term represents just a small correction to 
the main Wigner component. 
 
The same interrelation between the Wigner and Majorana components
should also be valid for the spin-orbit interaction and our stabilised
inversion tends to corroborate this conclusion. While the difference
in $J$ for the spin-orbit terms obtained from separate inversion of
even and odd partial waves is quite remarkable (see 
Fig.~\ref{fig-2ab}), this difference disappears for the potentials
obtained by the simultaneous inversion.  (Strictly speaking, it is
beyond the accuracy achievable with the existing PSA -- see the
Table~\ref{invs-cso}.)  Thus we may conclude that the initial
assumption on which the simultaneous inversion of even and odd parity
states was based, i.e. that the interaction is of mainly Wigner type,
is well justified by the results of the inversion.
  
\section{Inversion of RGM  $S(l)$ with breakup channel  
contributions} 
\label{rgm-sect} 
 
The resonating group method (RGM) offers fundamental advantages for
the study of the important effects contributing to $d + ^4$He
scattering, despite the simplified approximations necessary in the
method. Here contributions to the potential due to both the full
antisymmetrization of all nucleons and the inclusion of breakup
channels can be directly assessed in a way not possible by inversion
from empirical phase shifts.  This formulation of the $d + ^4$He RGM
calculations closely follows the work of Kanada {\em et al},
\cite{ad-dist}, and the breakup channels are included using the
pseudo-state method described therein.  Only $\Delta l =0$ transfer is
included in the breakup calculations.
 
The RGM $S$-matrices are numerically calculated using modified forms 
of the codes of Bl\"{u}ge {\em et al}, \cite{RGM-code}, adapted to 
incorporate the Minnesota $NN$ force, \cite{minn-pot} as described in 
a previous work, 
\cite{cc-calcs}. The exchange mixture parameter $u$ is set to $0.97$, 
but the potentials described below are not sensitive to this value. 
The deuteron and excited $n$-$p$ pseudo-states are described by an 8  
gaussian basis, \cite{kuk-Ryz}, which allows coupling to 4 distortion 
channels of low energy deuteron excitation. No spin-orbit force is 
included in these calculations since the empirical spin-orbit terms 
cannot be determined to sufficient accuracy. 
 
Tabulated $S$-matrix values are obtained from the RGM code for input
into the IP inversion and, as in Sect.~\ref{emp-sect}, the inversion
is confined to the energy range $0 < E_d < 15$ MeV.  With inclusion of
channel coupling the $S$-matrix becomes complex. At energies less than
10 MeV $|S|$ is close to unity, but at higher energies, the coupling
to breakup channels produces an irregular variation of $|S|$ with
energy. At the higher energies the imaginary potential arising from
breakup coupling has been shown to have a marked variation with
energy, \cite{ou-dhe}.  The imaginary potential cannot then be
established to any reliable accuracy over the energy range currently
considered and all inversions in this section establish real potential
components only.
 
The theoretical RGM phase shifts can be reproduced by inversion to a
far greater accuracy than the empirical $S(l)$. However the most
accurate inversions lead to irregularly shaped potentials due to an
inherent, but small, energy dependence associated with the underlying
nonlocality.  The energy dependence of $S(l)$ cannot be reproduced by
simply introducing an energy dependence in the potential magnitude,
but requires a variation in the {\em potential form\/} with energy.
Since the details of such a dependence are far beyond what can be
determined empirically, only smooth energy independent solutions are
now considered and these lead to an adequate fit to the phase shifts.
 
Solutions are obtained for both the two inversion methods described in 
Sect.~\ref{invs-sect} i.e. for separate inversion of even and odd 
$l$-values and for simultaneous inversion to determine $V_W$ and 
$V_M$.  The choice of method is less critical in the case of 
theoretical $S(l)$ and, with a small inversion basis, (2 functions per 
component) the two methods lead to very similar parity dependent 
potentials.  The simultaneous inversions also show that a reasonable 
fit to the RGM $S(l)$ is possible with a $V_W$ term alone, so that the 
addition of the $V_M$ component introduces only a second order 
correction. 
 
Fig~\ref{fig-1coop} shows the even and odd $l$ potentials for all four
cases, with and without breakup coupling and using the two inversion
approaches.  The components
$V_W$ and $V_M$ for the potentials obtained by simultaneous inversion
are illustrated in Fig.~\ref{fig-2coop} together with the RGM direct
potential obtained from the double folding model for the Minnesota NN
force. The breakup channels primarily change the even $l$ phase
shifts only, \cite{ad-dist}, so that, although the even potential {\em
increases} significantly in magnitude on inclusion of breakup
coupling, there is little change in the odd component.  
 
The restriction to energy-independent potentials leads to some
uncertainties in the inversion but some interesting results are
revealed. The introduction of coupling produces both an increase in
the Wigner potential magnitude at small radii and a small decrease at
$r \sim 3$ fm. However, at larger radii the solution inclusive of
breakup coupling is very close to the direct potential multiplied by
$N_{\rm DF}$=1.572 (see also Fig.~\ref{fig-1ab}).
 
The large changes in the even $l$ phase shifts due to breakup coupling
significantly decrease the parity dependence (i.e. the amplitude of
the Majorana force) for $r < 2$ fm.  At larger radii there is a
considerable agreement in the parity dependence for all solutions, but
the strength of the $V_M$ term barely exceeds 1 MeV.
 
We can now compare the RGM potentials with the empirical solution
presented in Sect.~\ref{emp-sect}, which is indicated as the dotted
line in Fig.~\ref{fig-2coop}.  The limitations of the RGM calculations
and its dependence on simple interactions, preclude the possibility of
a very exact agreement with empirical results.  Indeed the magnitude
of $V_M$ is too small at both larger radii and near the nuclear
centre, although the latter problem is improved by the addition of
breakup coupling. Significantly, the inclusion of breakup effects
brings the Majorana component closer to the empirical $V_M$ potential,
although the RGM potential is still too small in magnitude for $r<2$
fm. As a consequence, the net effective $d$-$^4$He RGM potential is
close to a conventional {\em local} parity-independent potential.
 
A poorer agreement is found between the RGM solutions and the
empirical potential resulting from separate inversion, particularly
for the Majorana component, which presents
further evidence in favour of the simultaneous inversion.

\section{Validity of double folding model for $d$-$^4$He and similar systems.} 
\label{dfm-sect} 
 
In the last two or three decades the DF-model has become extremely
popular for the description of optical scattering, not only for $p$,
$^3$He($^3$H), $^4$He etc., but also for such loosely bound particles
as $d$, $^{6,7}$Li or $^9$Be
\cite{bib1,bib9,bib11},\cite{bib5}-\cite{bib35}.  However, for the
latter projectiles the normalisation constant $N_{\rm DF}$ was found
to deviate considerably from unity and is anomalously low. This
problem has been explained
\cite{bib8,im-break,mk-break,bib5,bib6}, by the strong coupling
effects of the virtual or real breakup of the loose projectile.
 
The accurate and fairly reliable potential for $d$-$^4$He established 
by inversion now allows us to study the behaviour of the normalisation 
constant in a more quantitative manner than was possible in the simple 
phenomenological analyses of deuteron scattering by medium and heavy 
nuclei \cite{bib14}.  
 
The most important question here is how adequate is the DF potential 
form? The general success of the DF model apparently argues in favour 
of its correctness.  Nevertheless the situation here is far from 
trivial. In fact, the DF potential (which corresponds to the so called 
no-distortion approximation) must adequately describe only the 
peripheral part of inter-nuclear interaction where there is no 
distortion of the colliding nuclei.  Moreover, to achieve this it is 
necessary to use an $NN$-potential which has the correct peripheral 
behaviour, and the normalisation factor $N_{\rm DF}$ should be close 
to unity.  Then one hopes that the short range part of the 
inter-nuclear interaction, where the effects of distortions and 
breakup are of greater importance, will be screened by a strong 
imaginary potential. In such  
circumstances the DF model will be a good approximation for the exact  
interaction. However, most practical applications of the DF procedure  
are based on an effective {\em scalar} $NN$-potential (usually the 
M3Y-force) which is quite dissimilar to the true $NN$-force. The 
DF potential calculated in this way must then be renormalised with some  
constant $N_{\rm DF}\ne 1$. 
 
In such an approach the degree of similarity between the ``exact'' 
potential and the DF potentials in both the peripheral and short-range 
radial regions is very interesting.  Below we will try to answer these 
important questions. 
 
In the Fig.~\ref{fig-1ab}a a DF potential, calculated with the 
effective $NN$ Minnesota force, is compared with the stabilised 
solution of the inverse scattering problem described in 
Sect.~\ref{invs-sect}.  It is evident that the {\em initial} DF 
potential, even in the far peripheral region, is very different from 
the ``true'' interaction potential and the difference is rather large  
in the inner region. Now, following to the standard routine 
\cite{bib14}, we  can also  calculate the normalisation coefficient 
$N_{{\rm DF}}$ for which this DF potential best reproduces the phase 
shifts data.  The even-parity partial amplitudes are most accurately 
reproduced with $N_{{\rm DF}}$=1.572. 
 
The quality of fit to the phase shifts reached with this value of
$N_{{DF}}$ is displayed in Fig.~\ref{fig-6}.  Clearly, with the
exception of the $S$-waves, the renormalised DF-potential provides a
very reasonable fit to these phase shifts.  In fact, the direct
comparison of the renormalised DF-potential with potential obtained by
inversion (see Fig.~\ref{fig-1ab}a) reveals that the peripheral parts
($r\ge 3$ fm) of both potentials are in  reasonable agreement. In
the inner region, however, there are large differences between the
potentials and these differences are reflected in the $S$-wave phase
shift behaviour.  The $D$ wave and all higher partial phase shifts are
not influenced by these short range differences.  The role of the
short range region will be further reduced by the imaginary potential
which is present for $E_d>3.5$ MeV and which will screen out the short
range part of the potential.
 
Thus, the rather good agreement of both DF and inversion potentials at 
intermediate and large distances, and in the phase shift behaviour,
does support the  application of the DF procedure for the description 
of deuteron scattering, even by the very light $^4$He-nucleus. 
 
Further evidence is provided by a comparison of volume integrals, $J$,
for the various potentials, as shown in Table~\ref{invs-cso}. In fact,
the volume integral of the even parity potential, as determined by
simultaneous inversion, is very close (with a difference of $\sim 1\%
$ only) to that of the DF potential \cite{bib10}, calculated with the
standard (density-dependent) DDM3Y $NN$ effective force and with
$N_{{\rm DF}}$ adjusted to reproduce the even $l$ phase shifts.  The
equivalent comparison for the odd parity phase shift shows far less
agreement, and the volume integral of that M3Y DF potential for odd
$l$ is closer to the volume integrals for the even parity cases.  The
volume integrals for the DF potentials calculated with both the
Minnesota and M3Y-potentials and (after multiplication by appropriate
normalising factors $N_{\rm DF}$) are close to each other. However
the DF model with the M3Y force is expected to give a better agreement
with the data than the Minnesota force.

We can now turn to the role of deuteron breakup channels and
antisymmetrization effects. The basic problem here, as was emphasised
earlier, is how to reconcile the apparent contradiction between the
rather good fit to scattering data given by the double folding model
(with the M3Y force), on the one hand,  although highly
renormalised, \cite{bib9}-\cite{bib11},\cite{bib6},
\cite{bib32}-\cite{bib35}
and the proven contributions of the large effects of the virtual
deuteron breakup channels, \cite{bib11}-\cite{bib13}.  In fitting the
phase shifts with the DF model only the NN effective force constant is
adjusted, so that a good description of data means that the {\em
radial form} of deuteron polarisation potential\footnote {We assume
quite naturally that the excitation of tightly bound nucleus $^4$He at
$E_d\le$15 MeV can be safely ignored.} must be similar to the
potential giving rise to purely elastic scattering, i.e. without any
channel coupling. Thus the renormalised force constant in DF potential
should effectively incorporate three important effects:
\begin{enumerate} 
\item [(i)] coupling with virtual breakup channels; 
\item [(ii)] antisymmetrization effects; 
\begin{enumerate} 
\item[ ] and closely connected with the latter, 
\end{enumerate} 
\item[(iii)]  the odd-even splitting effect. 
\end{enumerate} 
 
Our conclusions derived here together with the results of other authors can be  
summarised as follows: 
\begin{enumerate} 
\item [-] the Majorana force component, although definitely contributing,  
plays a rather moderate ($\sim 8\% $) role and hence does not destroy 
the good quality of fit achieved with the pure central interaction 
potential of the DF approach; 
\item [-] antisymmetrization and channel coupling effects as   
discussed in Sect.~\ref{rgm-sect} of the present work and also in 
Refs.~\cite{bib12,bib13} are rather large when considered separately, but the 
effects compensate each other to a considerable degree\footnote {This 
is quite clear from a physical point of view since inclusion of the 
deuteron breakup components with a large radial range leads to an effective 
stretching of the deuteron while the antisymmetrization with the 
compact $\alpha $-cluster wavefunction acts in the opposite direction, 
i.e. produces a ``compressing'' effect on the deuteron .}. Thus, the 
resulting net effect is rather small (see Figs.~\ref{fig-1coop} 
and~\ref{fig-2coop}). 
\end{enumerate} 
 
The odd parity potential shows almost no change when the deuteron 
breakup channels are included, whereas the even potential changes, on 
average, by 15-25$\%$ at small radii.  The reason underlying this 
change is seen in Fig.~\ref{fig-2coop} where one can  see that both the 
Wigner term $V_W$ and the Majorana term $V_M$ deepen by $\sim $10 MeV 
near the origin on inclusion of channel coupling. However, the 
interaction in the even partial waves is described by the combination 
$V_W(r)+V_M(r)$ while that for the odd partial waves is governed by 
the combination $V_W(r)-V_M(r)$. As a result, the channel coupling 
(CC) leads to a deepening of the even potential by $\sim $20 MeV near 
origin and to almost no change in the odd potential component. Thus, 
on average, the net effect of channel coupling is about 10$\%$. 
 
Moreover, as we have already observed, the net effect of CC is mostly 
short ranged (see the upper part of Fig.~\ref{fig-1coop}).  Because 
this short range contribution is often screened by an imaginary 
potential and is effectively invisible, the long range contributions 
of CC can be simulated by renormalising the total NN force constant to 
reproduce the $d+A$ scattering cross sections. In fact, the 
one-channel effective $d$-$^4$He RGM potential (see the curves 1 and 2 
on the upper part of Fig.~\ref{fig-1coop}), which takes into account 
only the antisymmetrization effects but not the coupling to breakup 
channels, is very similar to the renormalised DF potential (see the 
dot-dashed curve on Fig.~\ref{fig-1ab}a). 
 
 While the DF potential, as initially calculated, is somewhat
 different in form from the ``true'' interaction, especially in the
 inner region, by renormalising NN constant this difference can be
 minimised. In other words, the renormalised DF potential reproduces
 the``true'' potential in the peripheral region ($r>$2.5 fm) quite
 well. However, the renormalised DF potential is a little deeper than
 the ``true'' one in the intermediate region 1.5 fm $<r<$2.5 fm and
 also more shallow by $\sim20\%$ than the true potential in the
 innermost region, $r<$1.5 fm. Consequently, when the DF model is
 applied to deuteron scattering off light nuclei we expect to obtain
 quite a reasonable description of the elastic scattering cross
 section, but with some deviations from the data for the vector and
 especially the tensor analysing powers, which are sensitive to
 variations of the odd $l$ phase shifts and to interference effects.
 
A careful comparison of the potentials determined from RGM phase
shifts calculated {\em without} channel coupling effects (solid lines
in the upper part of Fig.~\ref{fig-1coop}) with the renormalised
DF-potential $V_{DF}(r)$ (the dot-dashed line on Fig.~\ref{fig-1ab}a)
shows that the deviation of the DF potential from the true one is most
likely to arise due to deuteron breakup channel coupling\footnote
{This conclusion is in agreement with the results of analysis of the
scattering of other weakly bound projectiles like $^{6,7}$Li, $^9$Be
etc. \cite{bib6,bib38}.}.  Therefore one expects the DF model to give
a highly accurate description when applied to the description of
scattering of more dense and heavily polarised projectiles such as
$\alpha $ particles. Recent work, \cite{bib33}, confirms this
conclusion.

Our main result in this section for the scattering of a loosely bound
deuteron from the $^4$He core can be formulated as follows: If the
scattering is described by only a full antisymmetrization of the $d+
^4$He clusters without breakup contributions, then the local
equivalent potential must contain a large Majorana component and this
scattering {\em cannot} be described by a simple parity independent
optical potential. However, the addition of antisymmetrized breakup
channels to the no-distortion approximation results in a strong
reduction of the odd-even potential splitting and the resulting
unified potential can be approximated reasonably well by a
renormalised DF potential. In other words, only when both the
antisymmetrization and $d$ breakup contributions are considered
together, can use of the DF model be justified.
 
\section{Tensor $d$-$^4$He potential and the $^6$Li quadrupole moment} 
\label{tens-sect} 
 
The $d$+$^4$He tensor interaction has been neglected in the preceding
sections due to the lack of reliable mixing parameters in the present
PSA data .  On the other hand, a tensor potential is absolutely
necessary to describe the quadrupole moment of $^6$Li nucleus.
 
It is well known that the very small negative value of $^6$Li
quadrupole moment ($Q=-0.0644$ fm$^2$) cannot be explained by the
standard three-body $\alpha + 2$N model, \cite{bib36}, but 
probably originates through the coupling between the $D$-wave
component of the $\alpha$-particle in the closed channel $d+d\to
^4$He and the $D$-wave component of the $\alpha +d$ relative
motion. The negative value of $Q$ is associated directly with the
negative value of the asymptotic mixing constant $\eta _D = -0.0125$
\cite{bib37}, i.e. the sign of the $D$-component of the bound-state
wave function (in the asymptotic region) must be {\em opposite} to
that of the $S$-component. A $^6$Li wave function of this type has
been formulated, \cite{bib38} and this model does give the correct
sign of quadrupole moment.
         
Here we attempt to reconstruct a $d$--$^4$He potential which describes
the $^3S_1$ and $^3D_1$ low-energy phase shifts and the main
properties of the $^6$Li ground state, including the quadrupole
moment. Consequently the data input into the inversion includes the
binding energy and quadrupole moment of $^6$Li together with the phase
shifts for energies up to 11 MeV.  To retain stability in the
inversion, we have restricted the expansions of both central and
tensor parts of interactions to just one term:
\begin {equation} 
V(r)=V_c(r)+V_t(r)S_{12}, 
\end{equation} 
\begin {equation} 
V_c(r)=-V_0\exp (-\alpha r^2) 
\end{equation} 
\begin {equation} 
V_t(r)=-V_1\exp (-\beta r^2) 
\end{equation} 
 
Table~\ref{ten-pot} lists the parameters of two possible potentials
satisfying the above criteria.  The properties of the $^6$Li ground
state calculated from these potentials are presented in
Table~\ref{ten-vals}, together with the corresponding experimental
values.  The $^3S_1$ and $^3D_1$ phase shifts and the mixing parameter
$\varepsilon _1$, evaluated with the two potentials, are shown in
Figs.~\ref{fig-9} and ~\ref{fig-10} (solid lines are for variant A,
dashed line for variant B). The negative values of $\varepsilon_1$ can
be obtained only from narrow tensor potentials. The use of wider
potentials results in both a change of sign of $\varepsilon_1$ and a
change of the asymptotic mixing constant of the ground state
$\eta_D$. Thus the range of the $d$-$^4$He tensor potential is much
less than that of the central potential.  This finding provides
evidence for a strong $d$-exchange contribution to the tensor
$d$-$^4$He force
\cite{bib36}\footnote{ This effect should be compared with the NN 
one-pion-exchange forces where both central and tensor components have
the same range, $\mu^{-1}$ ($\mu$ is the pion mass).}  because the
range of the exchange force in the exchange mechanism should be rather
short due to the small r.m.s. radius of $^4$He and the large binding
energy in the channel $^4$He$\to d+d$.
 
\section{Conclusions.} 
\label{concls} 
 
In this work we have studied some important problems relating to the 
interaction potential underlying the scattering of composite particles 
such as $d$+$^4$He.  In order to establish the true interaction 
potential, we solved an inverse scattering problem using our effective 
method of linearised iterations.  The input for the inversions include 
both phase shifts calculated from RGM coupled channel calculations, 
which incorporate virtual and real breakup channels, and the results 
of a recent phase shift analysis. 
 
The success of the inversions presented here depended considerably on 
combining the very reliable even partial wave amplitudes with the 
``noisy'' odd $l$ values within the empirical data set.  The Majorana 
exchange interaction was {\em explicitly} evaluated in the inversion, 
and by this procedure the odd parity potential is determined with 
greater stability. The resulting empirical potentials then represent 
an improvement on previous $d$+$^4$He potentials for which the odd 
parity potential was unreliable and inaccurate.  
 
The new stabilised empirical potential can be applied to 
predict improved phase shifts for the odd partial waves. In fact, 
these predictions agree quite well with the RGM predictions,
but only after inclusion of 
the breakup channel coupling, both in the Wigner and the Majorana 
components of the potential.
 
In fact, both the solutions (i.e. obtained from the RGM phase shifts and from
the PSA data set) contain only a small magnitude for the Majorana
component. Significantly, the breakup effects only contribute
noticeably to the even parity potential and have little effect on the
odd parity component.  This odd-even splitting is then expected to be
of minor importance for other $d+A$ systems.
 
We have also compared these ``exact'' interaction potentials with the 
potentials calculated using the popular DF model based on two 
different effective NN forces: 
\begin{enumerate} 
\item [-] DDM3Y NN-force, and 
\item [-] Minnesota NN-force. 
\end{enumerate} 
This comparison shows that, despite the models producing apparently 
different $d$-$^4$He DF potentials, on the whole the general agreement 
is reasonable. However, the DDM3Y force, multiplied by an appropriate 
normalisation factor, leads to a better agreement with empirical 
potentials and provide a quite satisfactory reproduction of the 
``experimental'' phase shifts (see also 
\cite{bib14,bib10},\cite{bib32}-\cite{bib35}). Nevertheless more 
detailed effects such as the odd-even splitting etc. are not included
in the DF approach.
 
This agreement of the DF M3Y model with the empirical potential arises
as a result of the interference of several effects, none of which is
apparently taken into account in the DF model, notably
antisymmetrization and breakup channels coupling. These two
contributions mutually compensate each other in the relevant region of
configuration space to the extent  that the DF approach plus an appropriate
normalisation factor provides a reasonable approximation to the
required interaction potential. 

An analysis of $d+ ^{40}$Ca scattering at $E_d = 52$ MeV,
\cite{bib32}, lead to similar results to those presented here. In
particular the interrelation between the DF potential for $d+ ^{40}$Ca
and the ``exact optical potential'' extracted from the data by a model
independent analysis equates strongly to our findings for $d +
\alpha$. This comparison then suggests that our results may have a
much wider applicability than the particular case considered here.

Combining the above results leads to a very powerful inversion 
procedure.  The modified IP inversion method, as employed widely in 
the present work, converges very fast and in a stable manner with a 
good choice of initial potential. The DF model is now confirmed as a 
very convenient candidate for this initial approximation. The complete 
method, when invoked for inversion directly from cross-section data, 
\cite{bib19,p-ox16}, offers many opportunities to study fine details 
of the interactions of complicated systems such as
$^{24}$Mg+$^{112}$Sn, $^{12}$B+A etc. and even for unstable
radioactive projectiles scattered off stable targets (see
e.g. \cite{he8-b8}).
 
In this paper we have also determined a $d$-$^4$He tensor interaction.
The resulting potential is rather short ranged and is relatively large in
amplitude ($\sim 30-40$ MeV). As has been previously suggested,
\cite{bib36}, this interaction may arise due to the very specific
exchange effect in which the ``inner'' deuteron in the D state of the
$^4$He core is exchanged with the outer valence deuteron.  This
exchange mechanism can be applied to explain both the short range
character of the $d$-$^4$He tensor force, through the very small
r.m.s. radius of $^4$He, and the {\em negative\/} value of the $^6$Li
quadrupole moment.  The second result is consistent with the fact that
the s-wave and d-wave components of the total $^6$Li wavefunction,
when projected onto the $d$-$^4$He channel, have opposite signs in the
asymptotic region.  Thus the form and strength of the $d$-$^4$He
tensor force at both low and intermediate energies deserve detailed
study. The results presented here may be considered an initial step in
this interesting direction.

Many characteristic features of interaction of the $d$-$^4$He system have  
been established in this and   a series of preceding works  
\cite{bib1,bib7,bib20,bib18,bib19,bib26,bib36}.  
In these works important features such as the Pauli principle 
manifestation and the appearance of Pauli-forbidden states, the 
description of higher partial waves, channel coupling effects, the 
interrelation with supersymmetrical partner potentials and the general 
manifestation of dualism repulsion-attraction in composite particle 
interaction have been studied in detail. In particular we have shown, 
\cite{bib1,bib2}, that the deep attractive interaction potentials in 
the systems $^4$He+$^4$He, $^4$He+N etc. arise as a consequence of 
well localised Pauli forbidden states and the appropriate  
conditions of orthogonality for the scattering wavefunctions to these 
forbidden states.  In turn, the 
structure of the Pauli forbidden states is very closely interrelated 
to the shell model structure of the whole unified system 
\cite{bib3,bib7}. Other interesting effects have been found involving
the joint action of antisymmetrization and (virtual) breakup in
$^6$Li, \cite{bib12}. For this case calculations have established
that, while the inclusion of deuteron breakup channels {\em
diminishes} the $d+^4$He cluster probability in the full three body
$^6$Li wavefunction, the subsequent antisymmetrization of the $n + p +
^4$He wavefunction {\em increases} the $d+^4$He cluster
probability. Again both these important contributions produce opposing
effects.  Thus the present paper can be considered, in some sense, as
a concluding work in the long series of our studies for $d+^4$He and
similar cluster systems.
 
\section*{Acknowledgements} 
 
The authors are grateful to Prof. V.Neudatchin, Dr. R.S. Mackintosh 
and our colleagues at the Moscow Institute of Nuclear Research for useful 
discussions in the course of the work. 
 
\newpage  
 
 

\newpage 

\begin{figure} 
\caption[fig 1a and 1b] 
{(a) The central part of d-$^4$He potential for even partial waves
reconstructed with PSA input data up to 15 MeV.  The solid line is the
result of simultaneous inversion (present work) and the dashed line represents
the result of separate inversion,
\cite{bib20}. Also shown is the
potential obtained by separate inversion including energies up to 33
MeV, \cite{bib20} (dotted line), the DF potential with the Minnesota
NN force (triple dot-dashed line) and this DF potential multiplied by
$N_{{\rm DF}}=1.572$ (dot-dashed line).\\
 
(b) The central part of d-$^4$He potential for odd partial waves
reconstructed with PSA input data up to 15 MeV. Solid line is the
results of simultaneous inversion (present work), dashed line  is 
from a separate inversion of odd-parity P and F waves
\cite{bib20} and  the dotted line shows the inversion based on the 
P- wave phase  shifts only.}
\label{fig-1ab} 
\end{figure} 
 
\begin{figure} 
\caption[fig 2a and 2b] 
{(a) The d-$^4$He spin-orbit potential for even partial waves
reconstructed with PSA input data up to 15 MeV.  The notation is
as in Fig. 1a.
\\ 
 
(b) The d-$^4$He spin-orbit potential for odd partial waves 
reconstructed with PSA input data up to 15 MeV.  The notation is
as in Fig. 1b.} 
\label{fig-2ab} 
\end{figure} 
 
\begin{figure} 
\caption[fig 3] 
{The $S$- and $D$-wave phase shifts for the potential reconstructed  
by simultaneous inversion for energies up to 
15 MeV (solid lines). Dashed lines show the results for inversion of  
even waves only \cite{bib20}. PSA data are designated: $\bullet $ -- 
from \cite{bib26}, $\triangle $, 
{\msam \char 78}, {\msam \char 004}, {\msam \char 007} from 
\cite{bib25}  
for $S$- and  $D_J$-waves $(J=L-1,L,L+1)$ respectively.} 
\label{fig-3} 
\end{figure} 
 
\begin{figure} 
\caption[fig 4] 
{ Similar to the Fig. 3 but for the $P$-wave phase shifts.} 
\label{fig-4} 
\end{figure} 
 
\begin{figure} 
\caption[fig 5] 
{ Wigner $V_W$ (solid line) and Majorana $V_M$ (dashed line)  
parts of central potential reconstructed from PSA input data up to 15 MeV.} 
\label{fig-5} 
\end{figure}

\begin{figure} 
\caption[fig 1cooper] 
{Even (upper panel) and odd (lower panel) parity potentials determined 
from single channel RGM $S(l)$, by the two inversion approaches, 
separate (solid line) and simultaneous (dashed line), and from RGM 
$S(l)$ with breakup channel coupling by separate( dotted line) and 
simultaneous inversion (dot-dashed line).} 
\label{fig-1coop} 
\end{figure} 
 
\begin{figure} 
\caption[fig 2cooper] 
{The components $V_W$ (upper panel) and $V_M$ (lower panel) for 
simultaneous inversion from single channel RGM $S(l)$, (solid line) 
and from RGM $S(l)$ with breakup channel coupling (dashed line) 
compared with the empirical potential from simultaneous inversion 
(dotted line) and the RGM direct potential (dot-dashed line).} 
\label{fig-2coop} 
\end{figure} 
 
\begin{figure} 
\caption[fig 6] 
{Comparison  between the $S$-, $D$- and $G$-wave phase shifts for the  
double folding potential multiplied by $N_{{\rm DF}}=1.572$ (solid lines)  
and the ``average'' phase shifts calculated with central part of the  
reconstructed potential (dashed lines). ($G$-wave phase shifts are  
multiplied by 10 for  convenience.)} 
\label{fig-6} 
\end{figure}

\begin{figure} 
\caption[fig 7] 
{The theoretical eigen phase shifts $\delta _\alpha $ and $\delta
_\beta $ (corresponding to the $^3S_1$ and $^3D_1$ phase shifts in the
uncoupled channels) and the PSA-results. The solid and open triangles
display the PSA $\delta _\alpha $ and $\delta _\beta $
respectively. The solid and dashed lines  correspond to the tensor
potentials A and B respectively.}
\label{fig-9} 
\end{figure}

\begin{figure} 
\caption[fig 8] 
{The comparison between the theoretical and experimental values
of the tensor mixing parameter $\varepsilon _1$. The solid and dashed
lines  correspond to the tensor potentials A and B respectively.}
\label{fig-10} 
\end{figure}

 
 
\clearpage

\begin{table}[t] 
 
\caption{ Values of volume integrals -$(4\pi /8)\int V(r)r^2 dr$  
[Mev$\cdot$fm$^3$] for different $d$+$^4$He potentials} 
\label{invs-cso} 
\bigskip 
 
\begin{tabular}{|c|c|c|c|c|} 
\hline 
        & simultaneous & separate & DF with        & DF with \\  
        & inversion,   & inversion & Minnesota-force & DDM3Y \\ 
        & present work &  \cite{bib20} &  &  \cite{bib10}  \\ \hline 
even waves & 665.6     & 672.3     & 646.9  &  675.0  \\ \hline 
odd waves  & 546.3     & 573.3     &        &  614.0   \\ \hline 
spin-orbit & 13.8      & 19.8      &        &          \\ \hline 
\end{tabular} 
\end{table} 
 
\begin{table} 
\caption{Parameters of the tensor potential for $d$+$^4$He.} 
\label{ten-pot} 
\medskip 
 
\begin{tabular}{|c|c|c|c|c|}\hline 
variant& $V_0$ [MeV] & $\alpha$, [fm$^{-2}$]&$V_1$ [MeV] & $\beta$, 
[fm$^{-2}$] 
\\ \hline 
A & -71.979 & 0.20 & 27.0 & 1.12 \\ \hline 
B & -77.106 & 0.22 & 40.0 & 1.60 \\ \hline 
\end{tabular} 
\end{table} 
 
\begin{table} 
\caption{The properties of the $^6$Li ground state, calculated with  
the reconstructed tensor $d$+$^4$He potential} 
\label{ten-vals} 
\medskip 
 
\begin{tabular}{|c|c|c|c|}\hline 
Properties & Potential A & Potential B & Experimental values\\ \hline 
$E_b$, [MeV] & -1.4735 & 1.4735 & 1.4735 \\ \hline 
$R_r$, [fm]  & 2.60    & 2.56   & 2.56(5) \\ \hline 
$R_f$, [fm]  & 2.53    & 2.50   & 2.54(5) \\ \hline 
$Q$, [fm$^2$]& -0.064  & -0.064 & -0.0644(5) \\ \hline 
$C_0^0$      & 1.9     & 1.9    & 2.15(10) \\ \hline 
$\eta_D$     & -0.0115 & -0.0120& -0.0125(25)\\ \hline 
$\mu_d/\mu_0$ & 0.848   & 0.847  & 0.822 \\ \hline 
$P_D$, \%    & 1.59    & 1.78   &        \\ \hline 
\end{tabular} 
\end{table} 
 

\begin{references} 
 
\bibitem{bib1} V.I. Kukulin, V.G. Neudatchin, Yu.F. Smirnov, 
I.T. Obukhovsky {\em Clusters as Subsystems in Light Nuclei}. Vol. 3, In
the series {\em Clustering Phenomena in Nuclei}, (Vieweg \& S. Verlag,
Braunschweig/Wiesbaden, 1983).
\bibitem{bib34} B.Buck, C.B.Dover and J.P.Vary, Phys.Rev. C{\bf 11}, 1803 
(1975); 
B.Buck and A.A.Pilt, Nucl.Phys. {\bf A280}, 133 (1977). 
\bibitem{bib2} V.I. Kukulin, V.G. Neudatchin, A.A. Sakharuk and
V.N. Pomerantsev,  
Phys.Rev.C{\bf 45}, 1512 (1992);
Yad.Fiz. (Sov. J.Nucl.Phys.) {\bf 52}, 402 (1990);
Phys.Lett.{\bf B255},482 (1991). 
 
\bibitem{bib3} H. Horiuchi, Progr.Theor.Phys. {\bf 64}, 184 (1980). 
\bibitem{rgm-rep} Y.C. Tang, M. LeMere and D.R. Thomson, Phys. Rep.
{\bf 47}, 167 (1978).
\bibitem{bib30} S.G.Cooper, Phys. Rev. C{\bf 50}, 359 (1994).
\bibitem{bib7} L.D. Blokhintsev, V.I. Kukulin, D.A. Savin and A.A. Sakharuk  
 Yad.Fiz. (Sov. J. Nucl.Phys.) {\bf 53}, 693 (1991). 
\bibitem{feschb} H. Feshbach, Ann. Phys. {\bf 5}, 357, (1958).
\bibitem{bib8} S.A. Sofianos, K.C. Panada and  P.E. Hodgson, 
J.Phys.G: Nucl. Part. Phys.  {\bf 19}, 1929 (1993). 
\bibitem{bib9} G.R. Satchler, {\em Direct Nuclear Reactions} 
(Clarendon Press, Oxford, 1983). 
 
\bibitem{bib11} T.K. Roy and S. Mukherjee, J.Phys.G.: Nucl.Phys. {\bf 13},  
1239 (1987); G.H. Rawitscher, Phys.Rev. C{\bf 9}, 2210 (1974);
Nucl.Phys. {\bf A241}, 365 (1975)
  
\bibitem{bib12} R.G.Lovas, A.T. Kruppa, R. Beck and F. Dickman,
 Nucl.Phys. {\bf A474}, 451 (1987); 
K Varga and R.G.Lovas, Phys. Rev. C{\bf 43}, 1201 (1991); 
 Y. Fujiwara and Y.C. Tang, Phys. Rev. C{\bf 43}, 96 (1991); Few Body  
Systems, {\bf 12}, 21 (1992):
H.Kanada, T.Kaneko and Y.C.Tang, Phys. Rev. C{\bf 38}, 2013 (1988). 
 
\bibitem{bib13} Y. Yahiro and M. Kamimura, Progr.Theor.Phys. {\bf 65}, 
2046; 2051  (1981); 
 V.I. Kukulin and V.N. Pomerantsev,  Yad.Fiz. (Sov.J. Nucl.Phys.) 
{\bf 50}, 27  (1989).
\bibitem{bib14} J.Cook, Nucl.Phys. {\bf A382}, 61 (1982). 
\bibitem{bib15} G.R.Satchler, L,W. Owen, A.J. Elwyn, G.L. Morgan, and  
R.L. Walter,  Nucl.Phys. {\bf A112}, 1 (1968). 
\bibitem{bib16} S.G.Cooper and R.S.Mackintosh,  Phys. Rev.  C{\bf 54}, 
  3133 (1996). 
\bibitem{pa-emp}  S.G. Cooper and R.S. Mackintosh, Phys. Rev.  C{\bf 43}, 
 1001 (1991). 
\bibitem{pa-rgm} S.G. Cooper, R.S. Mackintosh, A. Cs\'ot\'o, and R.G. 
Lovas, Phys. Rev.C {\bf 50}, 1308 (1994). 
 \bibitem{im-break} A.A.Ioannides  and R.S.Mackintosh, Phys. Lett. {\bf 169B},
113 (1986).
\bibitem{bib20} V.I. Kukulin and V.N. Pomerantsev, Yad.Fiz.  
(Rus.J. Physics of Atomic Nuclei) {\bf 60}, 1228 (1997). 
\bibitem{m3y-rep} G.R.Satchler and W.G. Love, Phys. Rep. {\bf 55}, 183
(1979).
\bibitem{bib17} V.I. Kukulin, V.N. Pomerantsev and J. Hora\v{c}ek, Phys.Rev.  
A{\bf 42}, 2719 (1990). 
 
\bibitem{bib18} V.I. Kukulin and V.N. Pomerantsev, Yad.Fiz. 
(Sov.J. Nucl.Phys.) {\bf 51}, 376 (1990). 

\bibitem{bib10} P. Mohr, V. K\"{o}lle, S. Wilmes, U. Atzrott, G. Staudt,
J.W. Hammer, H. Krauss and H. Oberhummer,  Phys.Rev. C{\bf 50},  
1543 (1994); Z.Phys. {\bf A349}, 339 (1994). 
 
\bibitem{bib19} V.I. Kukulin, V.N. Pomerantsev and S.B. Zuev, Yad.Fiz.  
(Rus. J.Physics of Atomic Nuclei) {\bf 59}, 428 (1996). 
\bibitem{mk-break}  R.S.Mackintosh and A.M.Kobos, Phys. Lett. {\bf 116B},
95 (1982).

\bibitem{bib21} A. Adahchour, D. Baye and P. Descouvemont, Nucl.Phys.  
{\bf A579}, 305 (1994). 
\bibitem{p-ox16} S.G. Cooper, Nucl. Phys. {\bf A618}  87 (1997). 
\bibitem{pb-nnuc} R.S. Mackintosh and S.G. Cooper, J. Phys. G: Nucl. Part.  
Phys. {\bf 23}, 565 (1997). 
\bibitem{o16-o16} S.G. Cooper and R.S. Mackintosh, Nucl. Phys.  
{\bf A576}, {308} (1994).
\bibitem{11li} S.G. Cooper and R.S. Mackintosh, Nucl. Phys. 
{\bf A582},  {283} (1995) 
\bibitem{robson} B.A. Robson, {\em The theory of Polarisation Phenomena},
(Clarendon Press, Oxford, 1974).
\bibitem{bib25} B. Jenny, W. Gr\"uebler,V. K\"onig, P.A. Schmelzbach and
C. Schweizer, Nucl.Phys. {\bf A397}, 61 (1983).  
 
\bibitem{bib26} E.V. Kuznetsova and V.I. Kukulin Yad.Fiz. (Rus. 
J.Physics of Atomic  Nuclei), {\bf 60}, 528 (1997); 
 V.M. Krasnopol'sky, V.I. Kukulin, E.V. Kuznetsova,  J. Hora\v{c}ek
and N.M. Queen, Phys.Rev. C{\bf 43}, 822 (1991). 
\bibitem{bib31} Y.Koike, Progr. Theor.Phys. {\bf 59}, 87 (1978). 
\bibitem{ad-dist} H. Kanada, T. Kaneko, S. Saito and Y.C. Tang, Nucl. Phys. 
{\bf A444}, 209 (1985). 
\bibitem{RGM-code} G. Bl\"{u}ge, K. Langanke and H.-G. Reusch, ``Computational 
Nuclear Physics 2'', Ed. K. Langanke, J.A. Maruhn and S.E. Koonin 
(Springer-Verlag, New York) (1993). 
\bibitem{minn-pot} D. R. Thompson, M. LeMere and Y.C. Tang, Nucl. Phys. 
{\bf A286}, 53 (1977). 
\bibitem{cc-calcs} S.G. Cooper, OU preprint OUPD9707, 
Nucl. Phys. {\bf A} (to be published).
\bibitem{kuk-Ryz} V.I.Kukulin and G.G. Ryzhikh, Prog. Part. Nucl. Phys. 
{\bf 34}, (1995) 397;
V.I.Kukulin and V.N. Pomerantsev, Sov. J. Nucl. Phys. {\bf 50}, (1989) 17. 
\bibitem{ou-dhe} R.S. Mackintosh and S.G. Cooper, OU preprint OUPD9708, 
Nucl. Phys. {\bf A} (to be published).
 \bibitem{bib5} M. Kamal, V.T. Voronchev and V.I. Kukulin, VANT  
(Problems of At. Science and Technique), Ser. fusion. {\bf No 3}, 
 37 (1989);  
J.Phys.G: Nucl.Part.Phys. {\bf 18}, 379 (1992). 
\bibitem{bib6} C. Samanta, S. Ghosh, M. Lahiri, S. Ray and
S.R. Banerjee, Phys.Rev. C{\bf 45}, 1757 (1992);
 T.Sinha, S. Ray and C. Samanta, Phys.Rev. C{\bf 48}, 785 (1993). 
\bibitem{bib32} M.Ermer, H.Clement, P.Grabmayr, G.J.Wagner, L.Friedrich 
and E.Huttel,  Phys.Lett. {\bf B188}, 17 (1987). 
\bibitem{bib33} U.Atzrott, P. Mohr, H. Abele, C. Hillenmayer
and G. Staudt,  Phys.Rev. C{\bf 53}, 1336 (1996). 
\bibitem{bib35} H.Abele and G. Staudt, Phys.Rev. C{\bf 42}, 742  (1993).  
\bibitem{bib36}  V.I. Kukulin, V.N. Pomerantsev, Kh.D. Razikov  
V.T. Voronchev and   G.G. Ryzhikh, Nucl.Phys. {\bf A586}, 151 (1995).  
 
\bibitem{bib37} H.R.Weller and D.R.Lehman,  Ann. Rev. Nucl. Part. Sci. 
{\bf 38},  563 (1988);  
R.Crespo, A.M.Eiro and F.D.Santos, Phys.Rev. C{\bf 39}, 305 (1989). 
 
\bibitem{bib38} N.Nishioka, J.A.Tostevin, R.C.Johnson and K.I.Kubo,  
Nucl.Phys. {\bf A415}, 230 (1994). 
\bibitem{he8-b8} S.A. Goncharov, A.S. Dem'yanova and A.A. Ogloblin,
`` Nucleon density
distributions in $^8$He and $^8$B from elastic scattering by $^3$He and
 $^3$H'', Preprint INP MSU 97-31/4482, 1997, Moscow, Inst. Nucl. Phys.
Moscow State Univ.
 
\end{references}
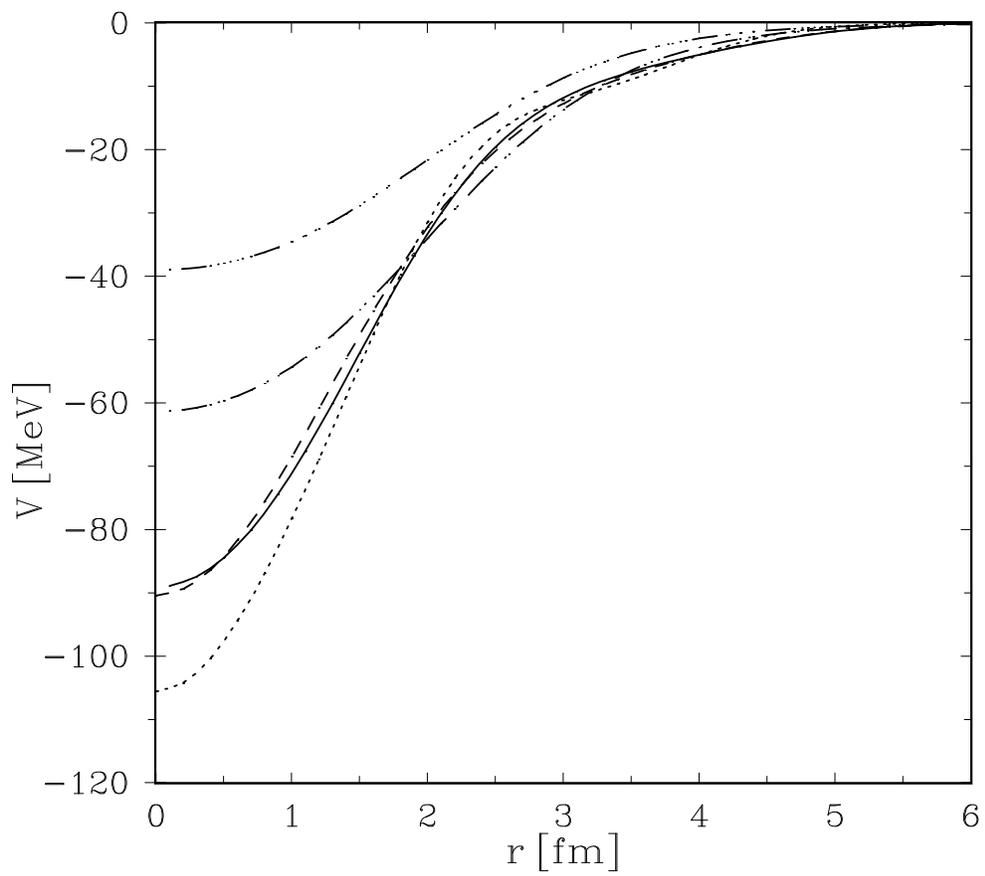
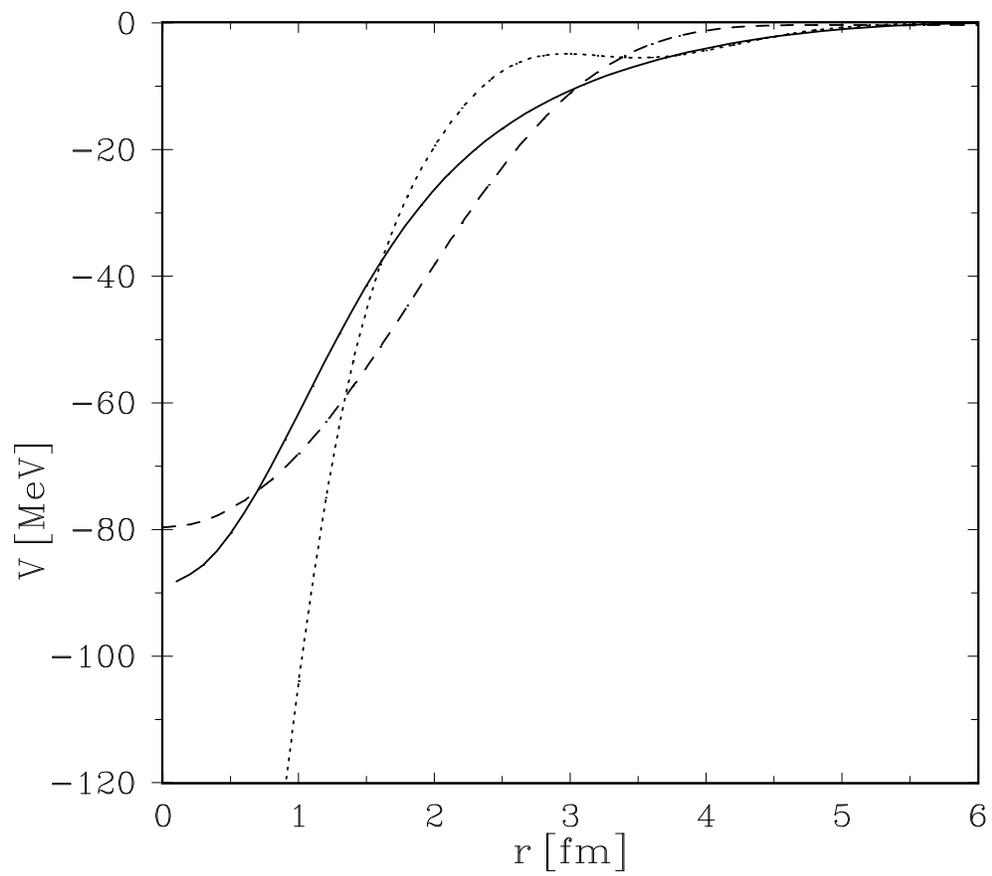
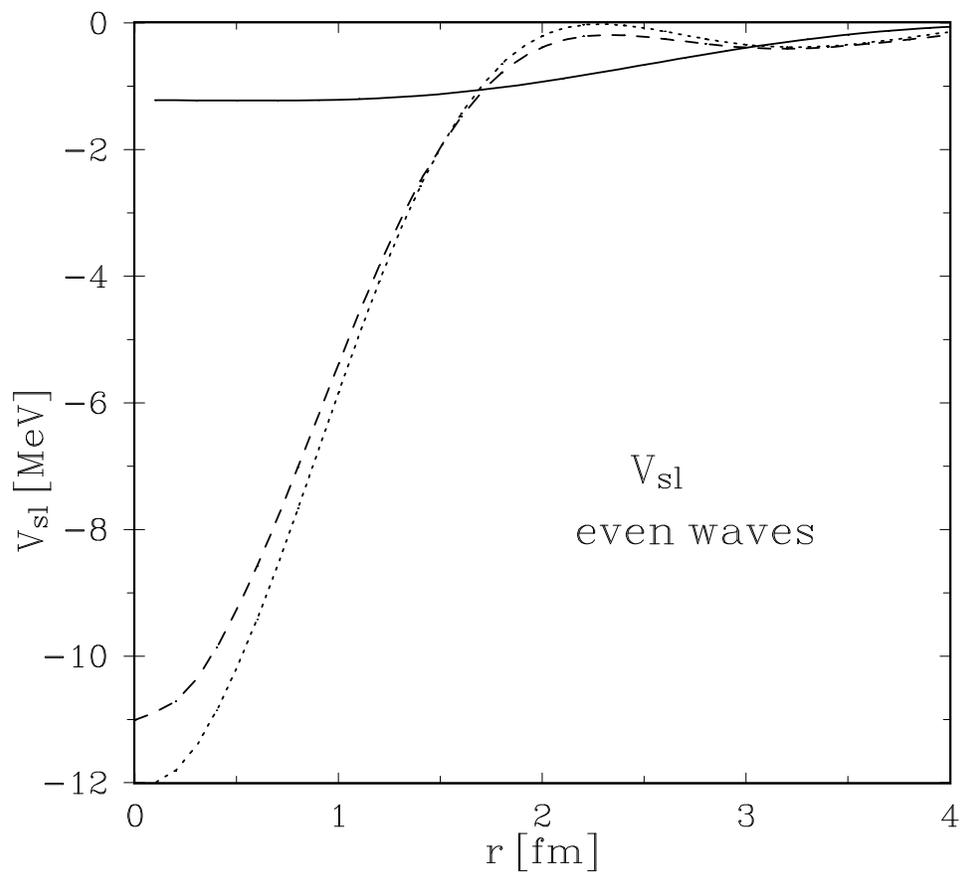
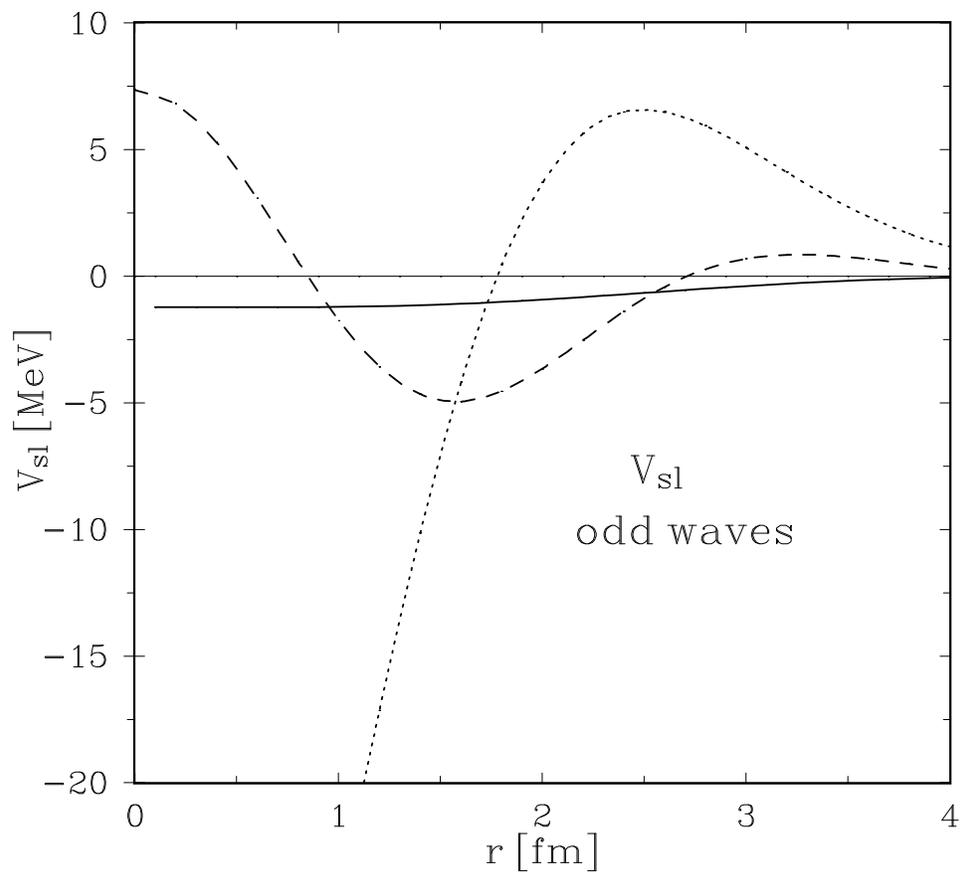
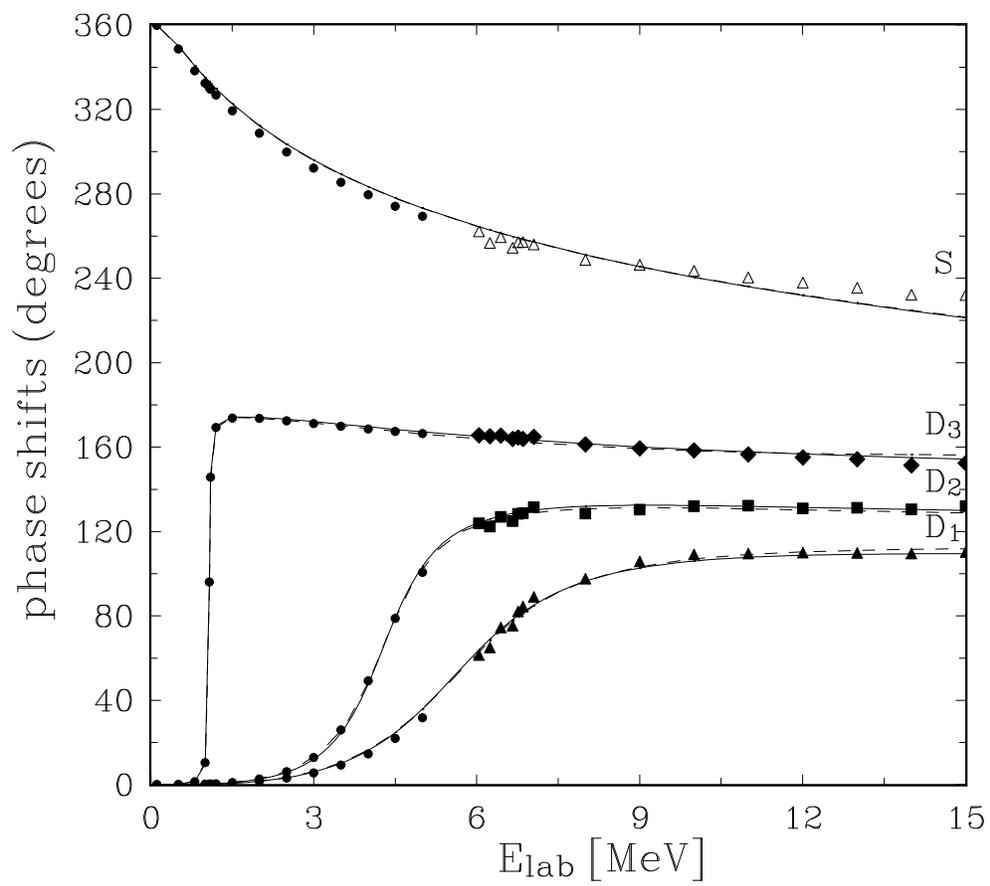
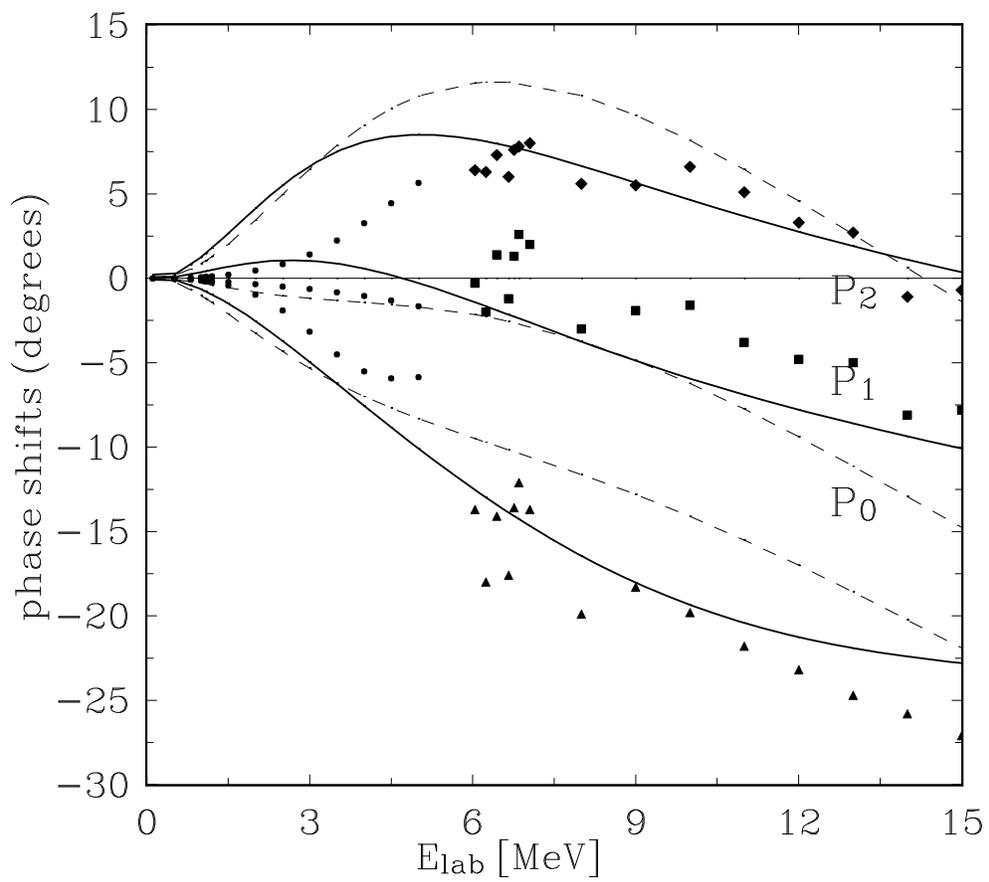
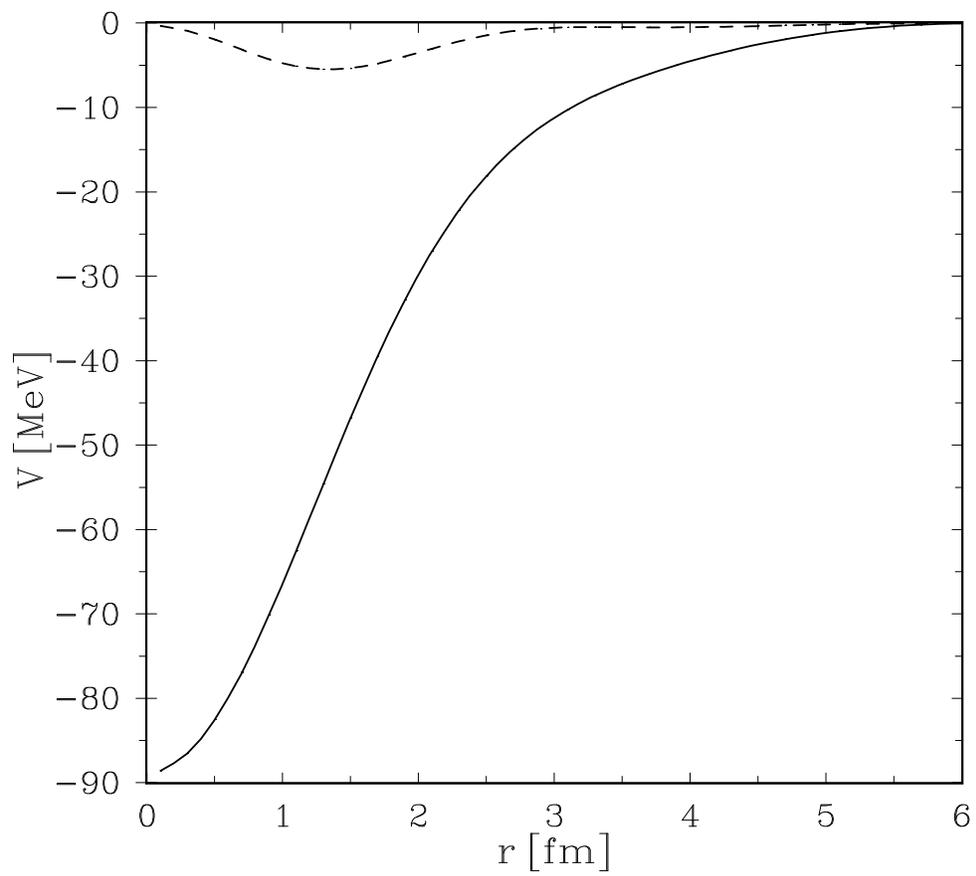
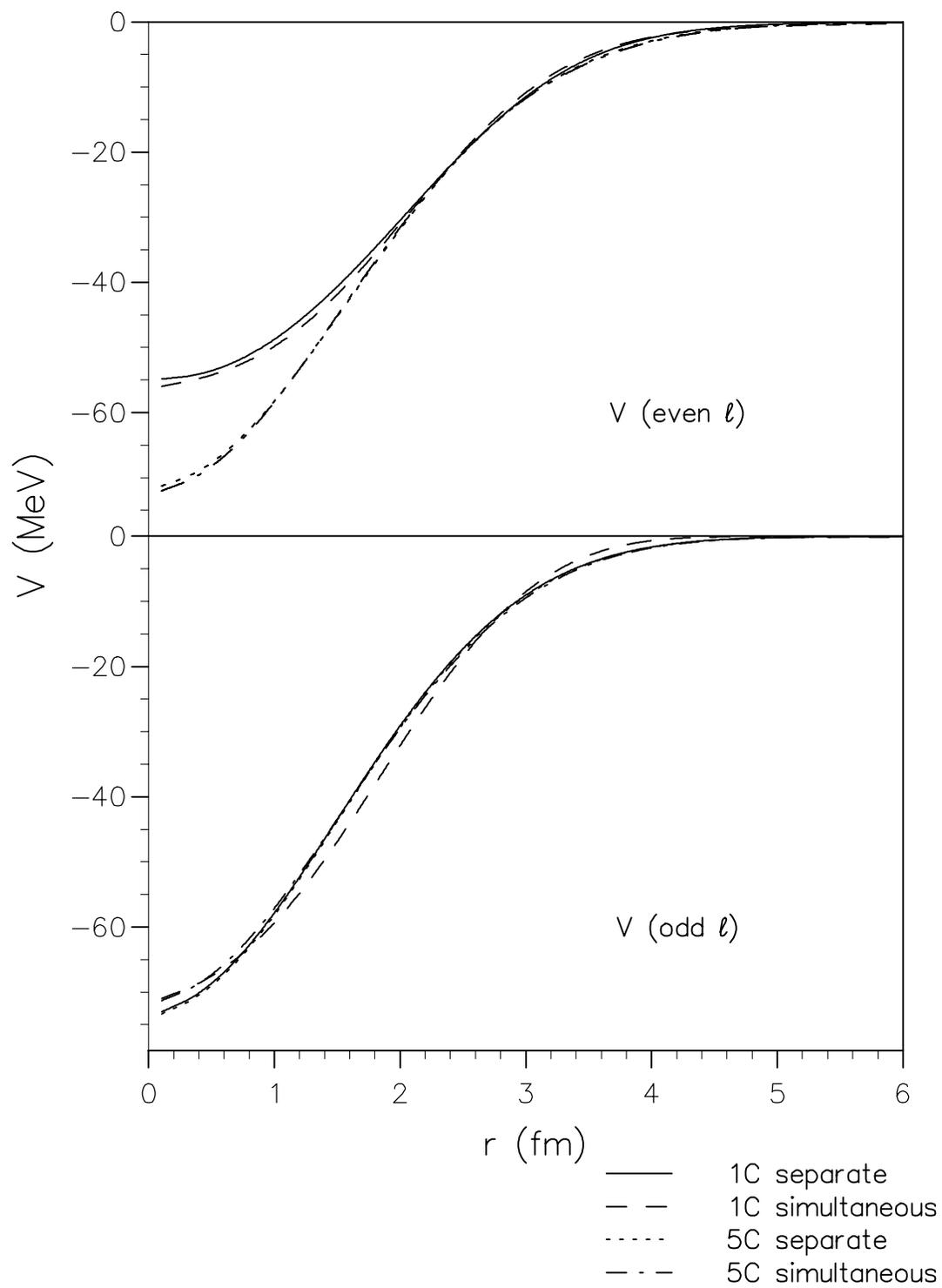
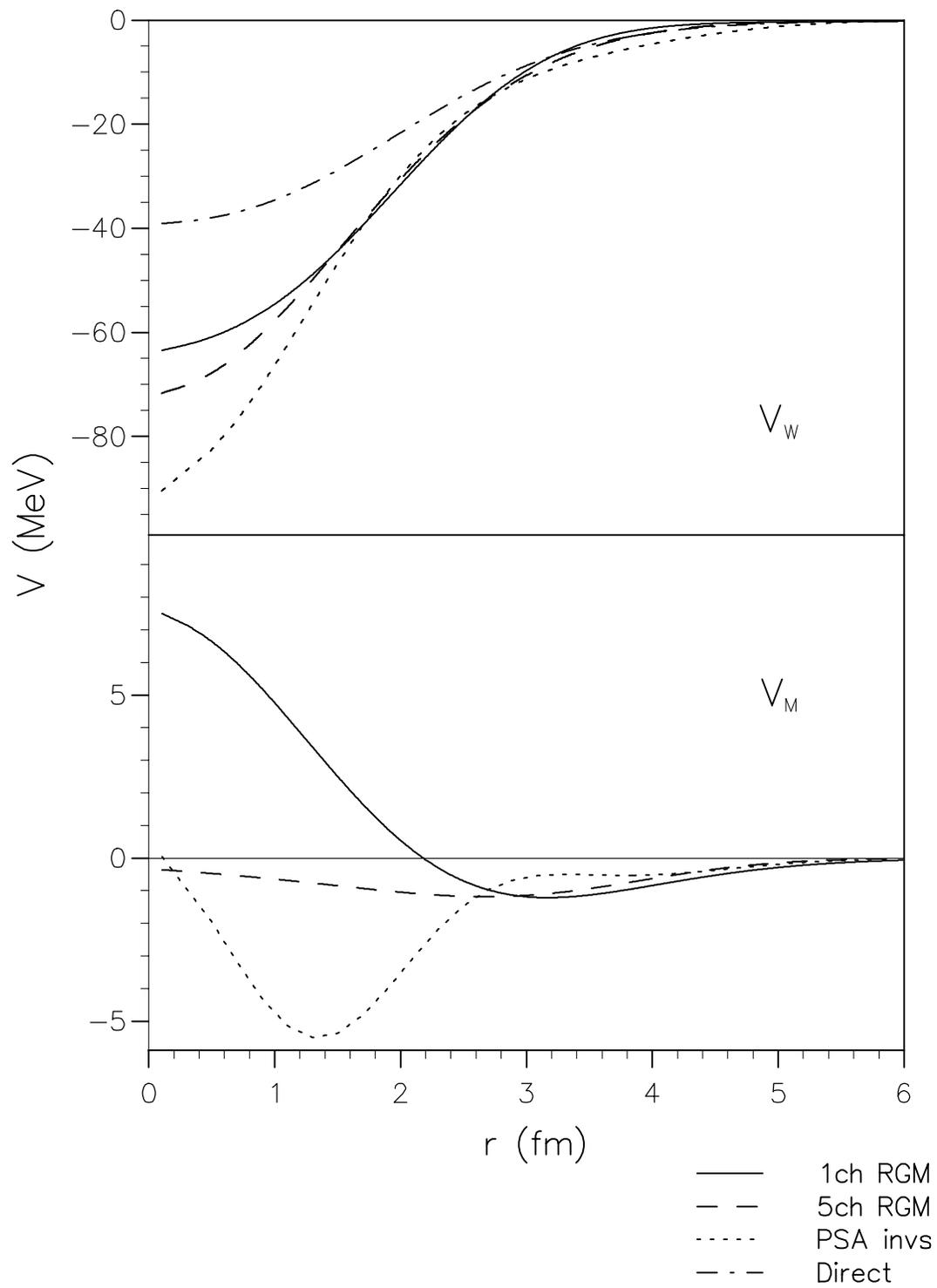
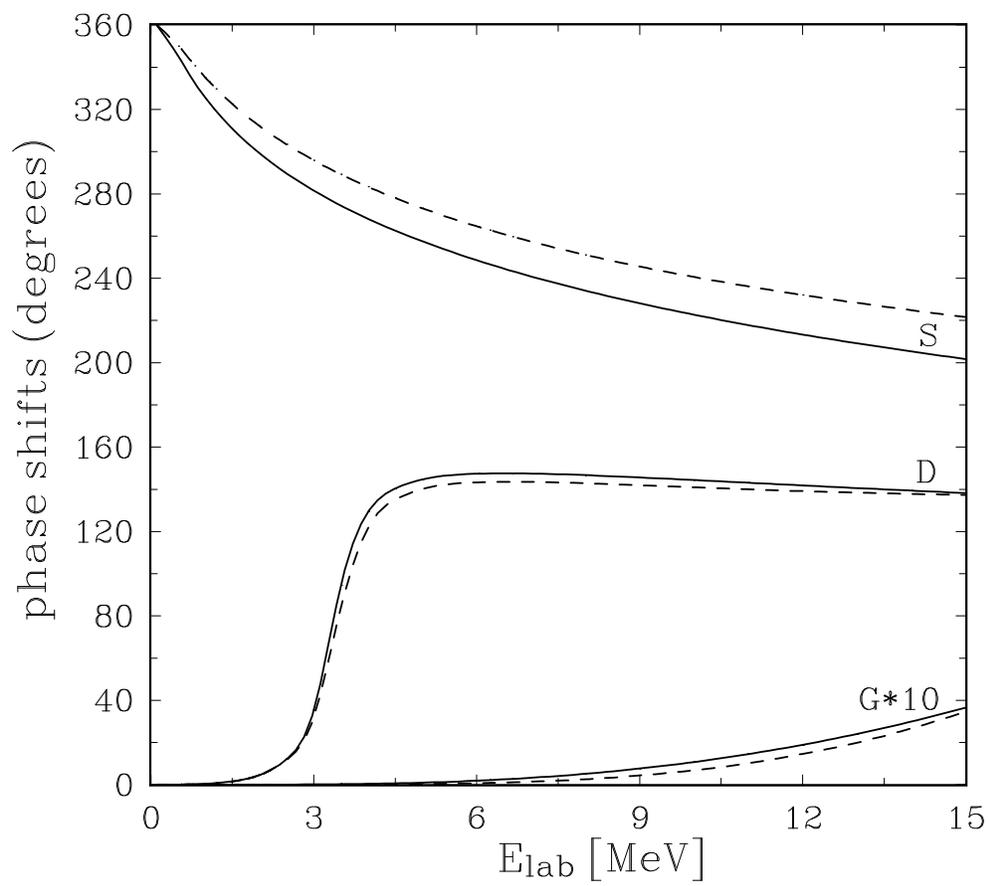
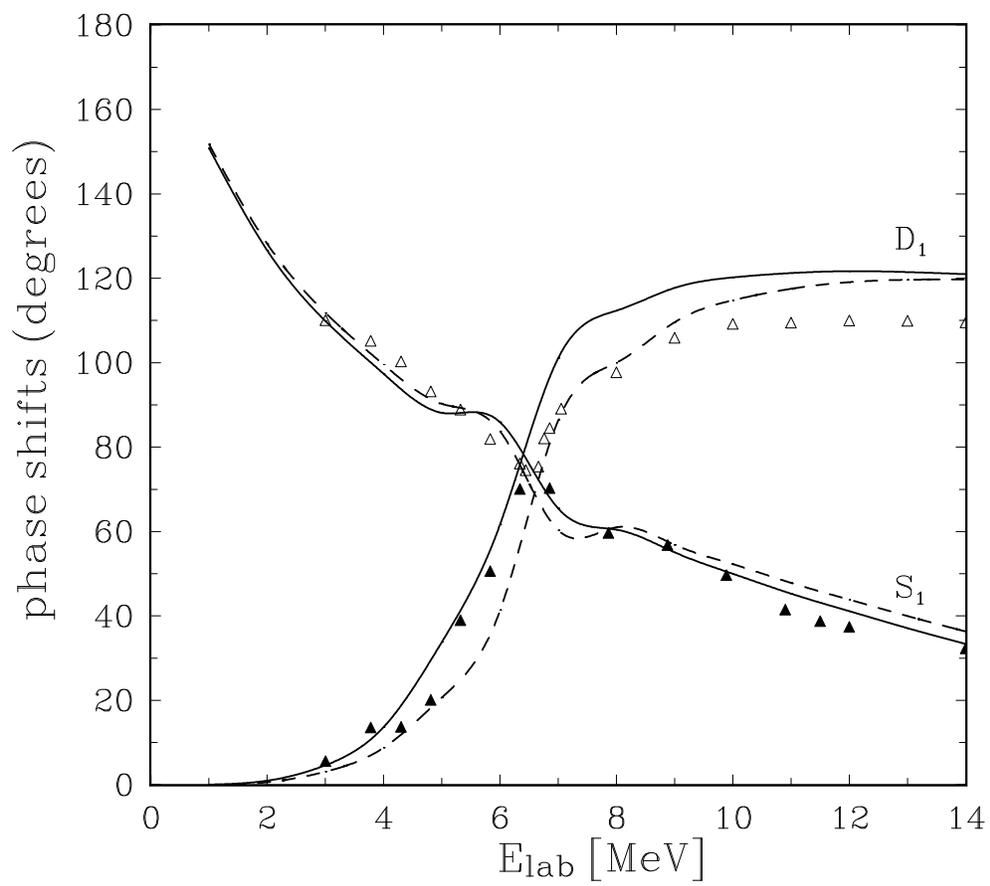
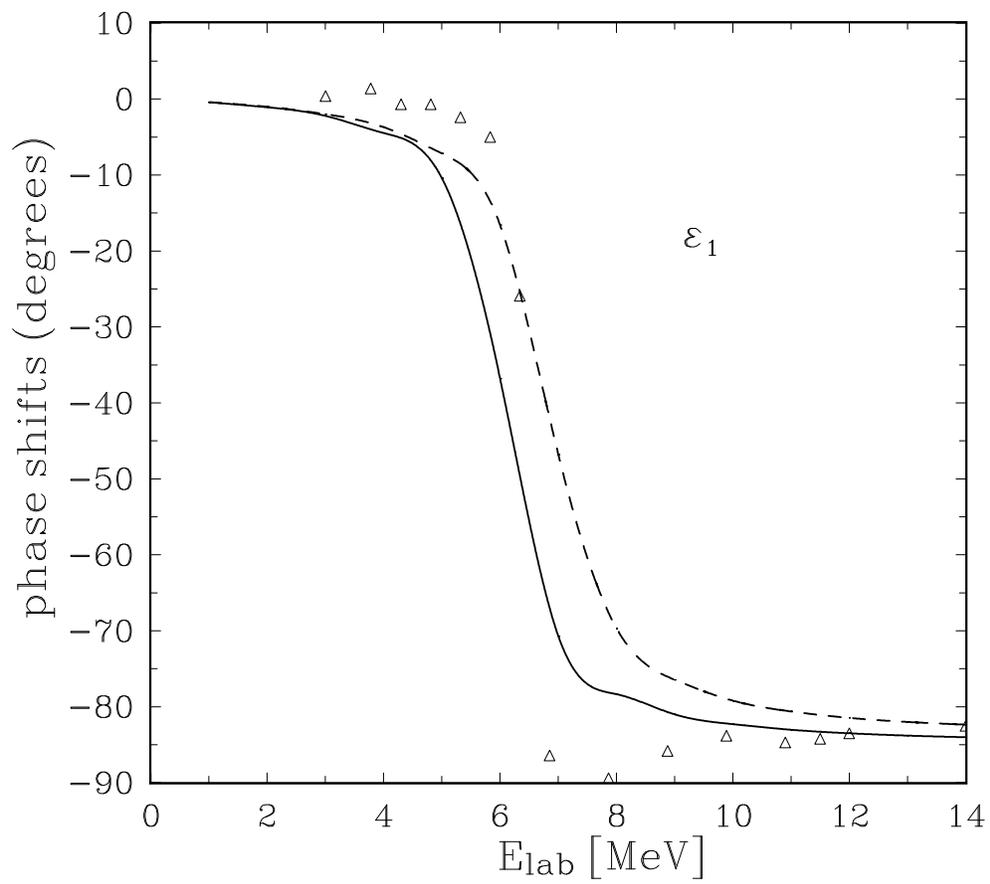
\end{document}